# Effects of mechanical stress, chemical potential, and coverage on hydrogen solubility during hydrogen-enhanced decohesion of ferritic steel grain boundaries: A first-principles study

Abril Azócar Guzmán [*] and Rebecca Janisch [†]

*Interdisciplinary Centre for Advanced Materials Simulation (ICAMS), Ruhr-Universität Bochum, 44801 Bochum, Germany*

Hydrogen-enhanced decohesion (HEDE) is one of the many mechanisms of hydrogen embrittlement, a phenomenon that severely impacts structural materials such as iron and iron alloys. Grain boundaries (GBs) play a critical role in this mechanism, where they can provide trapping sites or act as hydrogen diffusion pathways. The interaction of H with GBs and other crystallographic defects, and thus the solubility and distribution of H in the microstructure, depends on the concentration, chemical potential, and local stress. Therefore, for a quantitative assessment of HEDE, a generalized solution energy in conjunction with the cohesive strength as a function of hydrogen coverage is needed. In this paper, we carry out density functional theory calculations to investigate the influence of H on the decohesion of the $\Sigma5(310)[001]$ and $\Sigma3(112)[1\bar{1}0]$ symmetrical tilt GBs in bcc Fe, as examples for open and close-packed GB structures. A method to identify the segregation sites at the GB plane is proposed. The results indicate that at higher local concentrations, H leads to a significant reduction of the cohesive strength of the GB planes, significantly more pronounced at the $\Sigma5$ than at the $\Sigma3$ GB. Interestingly, at finite stress, the $\Sigma3$ GB becomes more favorable for H solution, as opposed to the case of zero stress, where the $\Sigma5$ GB is more attractive. This suggests that, under certain conditions, stresses in the microstructure can lead to a redistribution of H to the stronger grain boundary, which opens a path to designing H-resistant microstructures. To round up our study, we investigate the effects of typical alloying elements in ferritic steel, C, V, Cr, and Mn, on the solubility of H and the strength of the GBs.

## I. INTRODUCTION

Hydrogen embrittlement (HE) is a fundamental problem in materials science. In particular, it is known to have a detrimental effect on the mechanical properties of structural materials such as iron and iron alloys. For over a century, researchers have striven to understand the mechanisms of HE, still, many questions remain open [1]. This is mainly due to the fact that hydrogen changes the properties of several defects in the material, often at the same time. Common to all, however, is the adsorption of hydrogen in the first place. Therefore, it is important to understand and to be able to predict the solution, respectively trapping of H at vacancies, dislocations, grain boundaries, and in areas of residual strain in the microstructure or the stress field of crack tips. At the same time, there is the need to investigate the effects that H is causing at these defects. Only with a thorough understanding of these phenomena, multiscale mechanical models of hydrogen transport and embrittlement, and thus methods to prevent H embrittlement can be developed.

*Ab initio* density functional theory (DFT) calculations are a powerful tool to determine solution and trapping energies and interpret them in terms of the electronic structures. Several DFT studies have confirmed the tendency of H to segregate to grain boundaries and the effect of alloying elements thereon, e.g., [2–7], or investigated the trapping of H in vacancies in the bcc [8,9] and fcc Fe [10] lattice. The latter studies also discuss the maximum solubility of H at these defects. Even the solubility of H at dislocation cores can be estimated [11] and compared to other defects. However, the resulting partitioning should depend significantly on the H chemical potential and other factors, such as stresses in the microstructure. These two aspects have so far been neglected in first-principles studies of H segregation at grain boundaries.

In the study at hand, the focus is on grain boundaries (GBs), which play several roles in the context of mechanical properties of the material, even without hydrogen. They are known to have a significant impact on the deformability, strength, and fracture toughness of structural materials, such as iron and iron alloys. In hydrogen-charged systems, they provide trapping sites and thus remove mobile hydrogen from the grain interior, but they can also act as diffusion paths. Kirchheim [12] has emphasized the impact of excess hydrogen on the GB energy, as well as the chemical potential on metal systems [13]. Grain boundaries are expected to be prone to hydrogen-enhanced decohesion (HEDE), leading to intergranular fracture [14]. The HEDE mechanism has been attributed to weakening of interatomic bonds, due to the charge transfer between H and the host metal atom. There is no direct experimental evidence supporting this claim; however, the densities of state and electron densities obtained from electronic structure calculations of H at the $\Sigma 5(310)[001]$ STGB in bcc Fe [5] indicate charge redistribution from Fe to H at the GB. Again, electronic structure calculations represent a robust method to elucidate this mechanism further at the atomic and electronic scale.

*Present address: Institute for Advanced Simulations–Materials Data Science and Informatics (IAS-9), Forschungszentrum Jülich GmbH, 52425 Jülich, Germany; a.azocar.guzman@fz-juelich.de

†Contact author: rebecca.janisch@rub.de

The decohesion of cleavage planes in bcc Fe due to H presence was investigated by Katzarov and Paxton [15], where a reduction of the cohesive strength from 33 GPa to 22 GPa (approximately 33%) was reported. However, similar studies in symmetrical tilt grain boundaries (STGBs) did not find such a strong reduction, although HEDE is expected to occur at GB planes, promoting intergranular fracture. Tahir *et al.* [5] reported only a reduction of 6% at the $\Sigma 5(310)[001]$; while, Momida *et al.* [16] found a 4% reduction of the $\Sigma 3(112)[1\bar{1}0]$ ideal strength. The methodological differences between the techniques used to calculate the cohesive properties were addressed and reconciled in [17], where it is shown that if the excess elastic energy during the separation process is taken into account, there is a significant reduction of the GB strength by the presence of H, but it is not higher than that of the bulk (001) and (111) cleavage planes in the investigated range of H concentration at the GB.

This introduces the next aspect, since, as will be shown in the paper at hand, it is not only important to correctly determine the excess elastic energy during the separation of the grain boundary, but also to consider much higher local concentrations of H at the GB plane than what is usually studied in DFT calculations, when only one segregation site per structural unit is assumed for hydrogen atoms. In this paper, we propose a method to identify the initial configurations of H atoms at higher coverage of the GB, based on an algorithm that determines the voids in the atomic structure and the possible segregation sites. The solution energies then show that, similar to the vacancies [10], most GB structural units can capture several H atoms, leading to an even more pronounced reduction in strength.

To add to the complexity, it must be noted that the concentration-dependent solution energies also vary with a variation of the H chemical potential, and, most importantly for HEDE, they depend on the separation at the GB plane, i.e., the stress-dependent excess volume of the GB. In other words, the solution energy is a function of the reference H chemical potential, the local concentration, and the local stress. All these quantities can be coupled in a thermodynamic framework as introduced by Mishin [18] and Van der Ven and Ceder [19,20], but to the best of our knowledge has not been implemented for GBs so far, although it has been implemented for fracture studies of H in bulk metal [21].

Furthermore, Hirth and Rice [22,23] formalized the thermodynamic limits of the fracture process: (i) the limit of constant composition: the separation is faster and the interfaces may have empty segregation sites and (ii) the limit of constant chemical potential: a slower separation that occurs at a time scale that allows diffusion of the solute atoms to the interface. *In-situ* hydrogen-charged tensile tests of high-strength steels were performed by Depover *et al.* [24], finding that HE increased at lower deformation speeds. Further numerical methods elucidated that the stress dependency of the diffusion coefficient leads to different H concentration profiles [25]. In this study, we address both scenarios by calculating solution energies for several constant compositions, but also as a function of the H chemical potential for different coverage at the grain boundary and relating them to the local stress at the GB. This recipe on the one hand allows the prediction of the effective strength as a function of chemical potential by equilibrating the concentration during the separation process, as demonstrated for bulk cleavage planes by Katzarov *et al.* [15]. On the other hand, it can also be used to analyze the partitioning of H between different defects in the microstructure during mechanical loading, e.g., between the $\Sigma 5$ and $\Sigma 3$ STGB in this study. With this, a complete picture of the hydrogen distribution in a microstructure under various conditions can be obtained. This paves the way to identify the weakest links of a deformed microstructure and investigate ways to stabilize them. The latter point is addressed exemplarily by a study of the influence of the alloying elements C, V, Cr, and Mn in this paper.

## II. METHODOLOGY

### A. Technical details

We investigate the effect of the segregation of hydrogen together and without additional alloying elements to grain boundaries in bcc Fe. Total energies have been calculated using density functional theory (DFT) as implemented in the Vienna *Ab Initio* Simulation Package (VASP) [26,27], in a spin-polarized fashion. The electron exchange-correlation was estimated with the generalized gradient approximation (GGA) of the Perdew-Burke-Ernzerhof (PBE) [28] form and the core-valence interaction with the projector augmented-wave (PAW) method [29,30]. The convergence criteria for the electronic iteration was set to $10^{-6}$ eV and the equilibrium structures were relaxed until the forces on each atom were below $10^{-3}$ eV/Å. The $k$-point mesh was generated using Monkhorst-Pack grids [31].

Two symmetrical tilt grain boundaries (STGB) structures were chosen: $\Sigma 5(310)[001]\,36.9°$ and $\Sigma 3(112)[1\bar{1}0]\,70.53°$. The constructed supercells are shown in Figs. 1(a) and 1(b), respectively. The $\Sigma 5$ STGB supercell has 40 bcc Fe atoms, 20 atomic layers, and lattice vectors: $3a_0[310] \times 1.5a_0[\bar{1}30] \times a_0[001]$. The $\Sigma 3$ STGB has 96 bcc Fe atoms, 24 atomic layers, and lattice vectors: $4a_0[112] \times a_0[11\bar{1}] \times 2a_0[1\bar{1}0]$. The value of $a_0$ is 2.837 Å for Fe, in agreement with the literature [32]. For these cells of pure iron, the k-point grids consist of $2 \times 4 \times 8$ ($\Sigma 5$) and $4 \times 2 \times 4$ ($\Sigma 3$), and are scaled accordingly if the cell size is changed.

For the cases with C segregation, due to the long-range interaction of the C atoms, an additional cell was constructed with 30 atomic layers (60 bcc Fe atoms), extending the lattice vector in the [310] direction to $4.5a_0$. For both H and C cases, the supercell was doubled along the [001] direction in order to access lower concentrations.

The stable configurations of the GB supercells were obtained by optimization in the direction perpendicular to the GB plane to accommodate the excess length of the GB [33]; which is 0.31 Å for $\Sigma 5$ and 0.16 Å for $\Sigma 3$. The calculated GB energy for the $\Sigma 5$ GB is higher than for the $\Sigma 3$, 1.58 J/m$^2$ and 0.43 J/m$^2$, respectively. Similarly, the excess length is higher in the case of the $\Sigma 5$ (0.31 Å and 0.16 Å, respectively), which is attributed to the local atomic environment being more closed-packed at the $\Sigma 3$ GB plane.

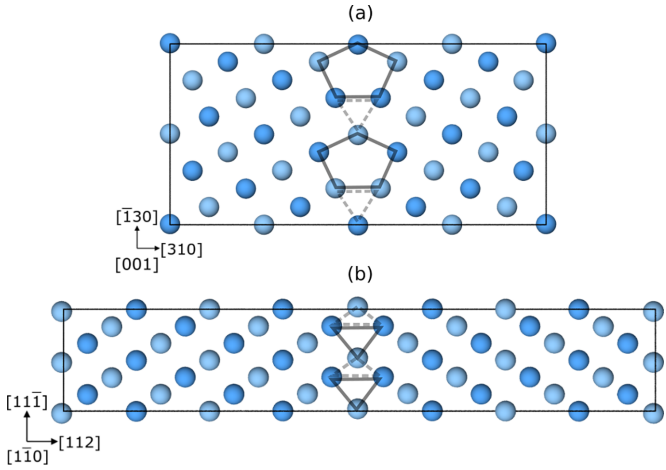

FIG. 1. Structure of the (a) $\Sigma 5(310)[001]$ and (b) $\Sigma 3(112)[1\bar{1}0]$ STGBs. The Fe atoms are indicated in blue, with the atoms of the structural unit of the GB delineated in grey.

### B. Solution energy

The solution energy $E_{sol}$ per interstitial atom can be obtained using the calculated total energies with different concentrations of H and C at the GB, as follows:

$$E_{sol} = \frac{E_{tot}^{GB+X} - E_{tot}^{GB} - \sum_{X=H, X=C} N_X \mu_X}{N_H + N_C}, \quad (1)$$

where $E_{tot}^{GB+X}$ is the total energy of the GB, either in pure iron, or with segregated substitutional elements Cr, V, or Mn, and containing $N_H + N_C$ segregated interstitials of type $X = H$ and/or $X = C$. Correspondingly, $E_{tot}^{GB}$ is the total energy of the GB, either in pure Fe or with segregated substitutional elements Cr, V, or Mn, but without any interstitial elements. The chemical potential of H or C $\mu_X$ refers to the reference value of H in a H$_2$ molecule (–3.385 eV) or C in diamond (–9.120 eV). Finally, $N_X$ is the number of interstitial atoms in the supercell.

### C. Ab initio tensile tests and the first principles cohesive zone model

A full characterisation of the GB cohesion requires, besides the work of separation, also the calculation of the tensile strength of the interface [34]. The decohesion process can be studied through the so-called *ab initio* tensile test. Although it has been extensively used for calculating the cohesive strength of material systems [35], there are certain intricacies that require careful implementation. The different approaches are described and compared in detail in [17].

One way to perform such a test is to introduce a certain displacement $\Delta$ in the supercell between the two grains defining the GB plane (or the bulk cleavage plane) and calculating the energy E, while keeping the positions of the atoms fixed. The energy-displacement data can be fitted using the universal binding energy relationship (UBER) [36] and the ideal cohesive or fracture stress $\sigma_{coh}$ can be obtained from $\sigma_{coh} = \partial E / A \partial \Delta$, where A is the area perpendicular to the cleavage plane. The maximum value of the stress corresponds to the cohesive strength.

The case of the rigid tensile test describes ideal brittle cleavage under loading mode I. However, the alternative approach, a tensile test in which the relaxation of atomic positions is allowed, is required to correctly predict segregation sites under stress and understand the effect of segregating atoms on the structural changes. In this case, due to the release of elastic energy, the output scales with system size, as shown by Nguyen and Ortiz [37] and Hayes *et al.* [38]. A solution was proposed by Van der Ven and Ceder [19,20], the first-principles cohesive zone model, which allows the derivation of a traction-separation law independent of the size of the system by using not the total, but excess energies. This approach has been extended to systems with GB planes and it is implemented in the present paper, the detailed description of the method can be found in [17].

Two different types of calculations are required for this excess energy approach: rigid grain shifts with subsequent relaxation (RGSrel) of the GB cell and a homogeneous elongation of a single crystal (HEC) in the same orientation as the grain boundary cell. The latter serves to determine the stress as a function of interplanar spacing in the bulk. This information can then be used to identify the stress in a GB supercell. The excess energy and length are calculated through the difference between RGSrel and HEC at the same stress. In the case of the cohesive zone containing a segregated GB plane, the excess energy and length with a defect (impurity and GB), $e_{ex}^D$ and $l_{ex}^D$, are calculated as follows:

$$e_{ex}^D(\sigma) = \frac{E_{RGSrel}^{GB+X}(\sigma)}{A} - (n_p - 1)\frac{E_{HEC}^{bulk}(\sigma)}{A} - N_X \mu_X, \quad (2)$$

$$l_{ex}^D(\sigma) = \tfrac{1}{2}\left[L_{RGSrel}^{GB+X}(\sigma) - (n_p - 1)L_{HEC}^{Fe}(\sigma)\right] - \Delta L^{GB}. \quad (3)$$

$E_{RGSrel}^{GB+X}$ is the total energy of the GB supercell with $N$ interstitial segregants, obtained from the RGS with relaxation. $E_{HEC}^{bulk}$ is the total energy *per atomic plane* from the HEC pure Fe bulk calculation. The number of atomic planes perpendicular to the tensile axis is denoted by $n_p$ and $A$ is the area projected onto the cohesive zone. Both energies are at the same stress $\sigma$ obtained via the interplanar spacing. In equation (3), $L_{RGSrel}^{FeGB+X}$ is the total length of the GB supercell in the direction of elongation and $L_{HEC}$ is the spacing between two adjacent planes in the HEC pure Fe bulk case. $\Delta L^{GB}$ is the excess length of the pure Fe grain boundary at zero stress.

Finally, the opening of the cohesive zone $\delta$ can be obtained from $\delta = l_{ex} - d_0$, with $d_0$ being the equilibrium interplanar distance at zero stress. The excess energy vs the opening of the cohesive zone can again be fitted to an UBER curve, and the derivative corresponds to the theoretical cohesive stress, as per the following equation:

$$\sigma_{coh} = \frac{1}{A}\frac{\partial e_{ex}^D}{\partial \delta}. \quad (4)$$

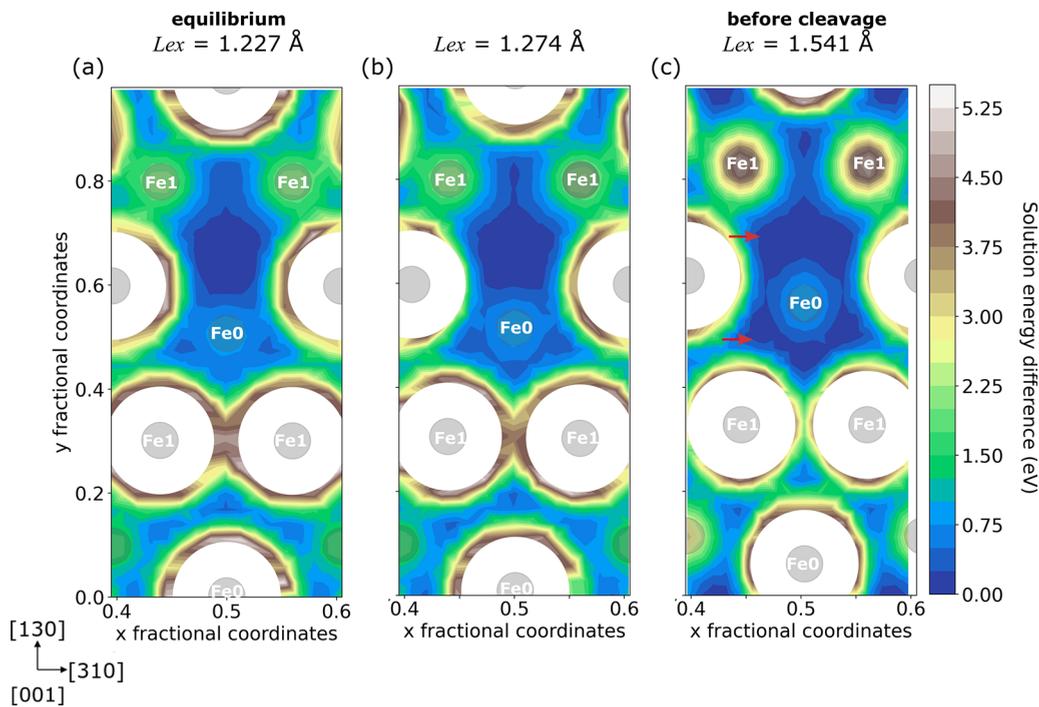

FIG. 2. Cross-section of the solution energy surface of the structural unit of Σ5 STGB in the Fe+H system. The H atom is placed in a range of ±2 Å from the GB plane in the $x$ direction. The H atom is placed every 0.25 Å in $x = [310]$ and $y = [\bar{1}30]$ directions. The [001] direction is kept constant at $z = 0$. The H atom was placed at a distance no closer than 1 Å from the in-plane Fe atoms.

The difference between the definition of excess length in Eq. (3) and the original one in [19] is the additional $\Delta L^{GB}$, which defines the opening of the cohesive zone as zero if the GB is free of segregated atoms and fully relaxed. Without this addition, even the pure GB would have a finite opening, even if it was stress free. However, this is a mere convention and does not change the derivative, i.e., the cohesive stress vs opening of the cohesive zone.

In all tensile tests performed in the present paper, the Poisson contractions are suppressed, which means uniaxial strain loading. From this, a traction-separation law for continuum fracture simulations under mode I loading and plane strain conditions can be derived. Under such a fracture mode, a triaxial state at the crack tip occurs [39], equivalent to the one obtained with the tensile tests of this paper.

### D. Identifying segregation sites

One of the open questions of H embrittlement mechanisms is how to predict the local concentration of H at different defects, specifically grain boundaries in the case of HEDE; and how the local concentration of H changes under mechanical load. To identify the possible segregating sites for H atoms at the Σ5 STGB, we have calculated the solution energy surface of one H atom in and in the vicinity of the structural units that define the GB, at equilibrium and under strain. The solution energy surface is obtained as the difference between the rigid energy of the system with H placed anywhere in the selected volume and the total energy of the most favorable configuration of 1 H atom at the GB.

Fig. 2 shows the solution energy surface in a plane perpendicular to the grain boundary at $z = 0$ with excess lengths corresponding to 0, 3, and 10% elongation. In the equilibrium GB [Fig. 2(a)], the minimum is located in the center of the trigonal prism created by the Σ5 structural unit, which is in agreement with the selected sites for H in this system found in the literature [5–7,16,17]. The most energetically favorable region is extended as the GB is strained up to 3% [Fig. 2(b)]. However, when higher values of strain are reached [Fig. 2(c)], the minimum separates into two distinct regions and more minima emerge around the central Fe atom (highlighted by red arrows). This means more H segregating sites are available in the Σ5 STGB as it expands. These sites can be identified through such energy surface calculations, but this approach is extremely time consuming.

An alternative method for identifying the segregating sites is proposed based on the algorithm of a polyhedral unit model by Banadaki and Patala [40]. This algorithm identifies fcc GB atomic structures by creating a three-dimensional array of polyhedra from the Voronoi vertices present in the structure, using a clustering technique it is possible to classify the geometries of the observed GB polyhedral units by comparing them with a database of hard-sphere packings. Here, we have implemented this algorithm for bcc lattices and analysed the individual voids in the structure available for segregation. The Python implementation of the algorithm is based on the Voro++ [41] and pyscal [42] software, the void analysis code is available in a public repository [43]. The identified voids for the segregation of atoms at the GB are compared with the most favorable positions found with the energy surface calculations. At 10% elongation, in both Figs. 2(c) and 3(a), comparable results from both techniques are obtained: Two distinct minima can be observed inside the trigonal prism of the structural unit and several smaller sites around the center Fe atom. Yet, the void analysis from the polyhedral unit model is substantially more efficient.

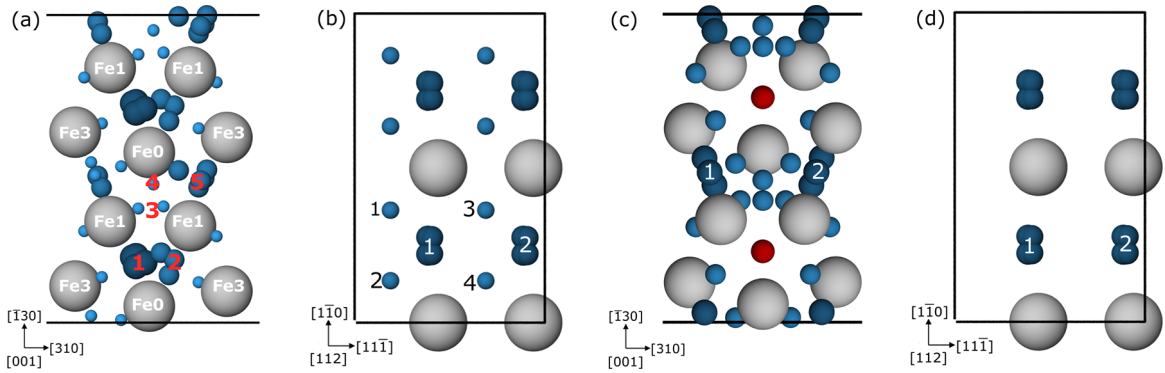

FIG. 3. Void analysis at 10% elongation of the pure Fe GB [(a),(b)] and the GB with segregated C atoms [(c),(d)]. Voids are identified with blue color, Fe atoms are colored grey and C atoms red. The Σ5 Fe STGB corresponds to (a) and (c) and Σ3 Fe STGB (b) and (d).

The concentrations of H and C chosen for the decohesion study were selected from the numbered sites in Figs. 3(a), 3(b) and 3(c), 3(d), respectively. For Σ5 STGB: $Fe_{80}H_2$, $Fe_{80}H_4$, $Fe_{80}H_6$, and $Fe_{40}H_4$ with only 1 H per structural unit; $Fe_{40}H_6$ as a mixed case and increasing up to four H atoms per unit $Fe_{40}H_8$, $Fe_{40}H_{12}$, and $Fe_{40}H_{16}$. Since C is a bigger atom, the void analysis was performed with an already segregated C at the GB [Fig. 3(b)] and the chosen concentrations are: $Fe_{120}H_2$, $Fe_{120}H_4$, $Fe_{120}H_6$, $Fe_{60}H_4$, and $Fe_{60}H_8$. And finally for Σ3 STGB: $Fe_{96}H_8$, $Fe_{96}H_{16}$, $Fe_{96}C_4$, and $Fe_{96}C_8$. These concentrations are translated to a GB coverage value $\theta$ calculated as the ratio of H or C atoms per Fe at the GB. It can be observed in Figs. 1(c) and 1(d) that the GB layers can be identified from the interlayer spacing deviation from the grain interior; from this, the Fe GB atoms are selected as six atoms for Σ5 and four atoms for Σ3.

## III. RESULTS

### A. Decohesion at constant concentrations of H

Initially, the tensile test is carried out for several fixed concentrations of H at the GBs. This is done by stepwise placing a H atom in each structural unit, until all units are occupied by one atom, then increasing the number of H per structural unit. The sites were chosen according to the void analysis described in Sec. II D. The coverage of the GB with solute atoms then is the ratio of the solute atoms per GB Fe atom, where GB atoms are those that define the structural units. The corresponding area concentrations and number of atoms per structural unit are found in Appendix A.

The excess energy and opening of cohesive zone calculated from Eqs. (2) and (3) are plotted in Fig. 4(a). The work of separation ($W_{sep}$) is significantly reduced with increasing H coverage; from 3.4 J/m² to 1.2 J/m² at 1.33 coverage. From the UBERfit of the excess energy curves the cohesive stress is calculated and shown in Fig. 4(b). Similarly to the work of separation, the cohesive strength is significantly reduced with increasing H coverage, from 20.6 GPa to 8.2 GPa at 1.33 coverage.

The same recipe was applied to the Σ3 STGB. The resulting excess energy and cohesive stress curves as a function of the opening of the cohesive zone are shown in Appendix B.

From the stress as a function of opening, one can obtain the maximum, which is the cohesive strength of the grain boundary for this particular coverage. The cohesive strength as a function of coverage is shown in Fig. 5 for both grain boundaries. For comparison, the equivalent results for the completely rigid scheme are shown as well, where the ideal cohesive stress is calculated according to Eq. (4).

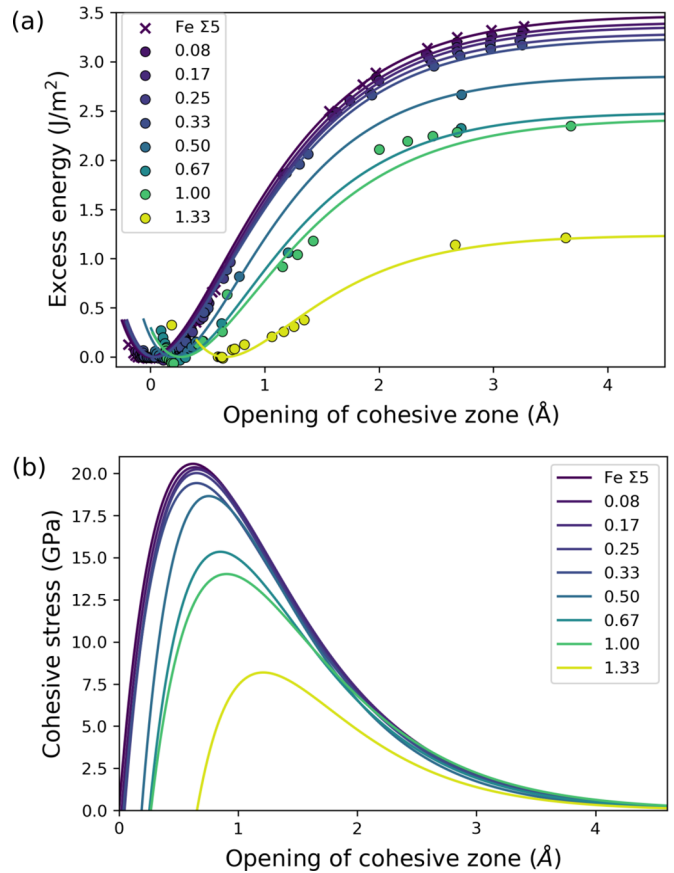

FIG. 4. (a) Excess energy and (b) cohesive stress as a function of the opening of the cohesive zone for the Σ5 STGB with varying H coverage (ratio of H/Fe atoms) at the GB.

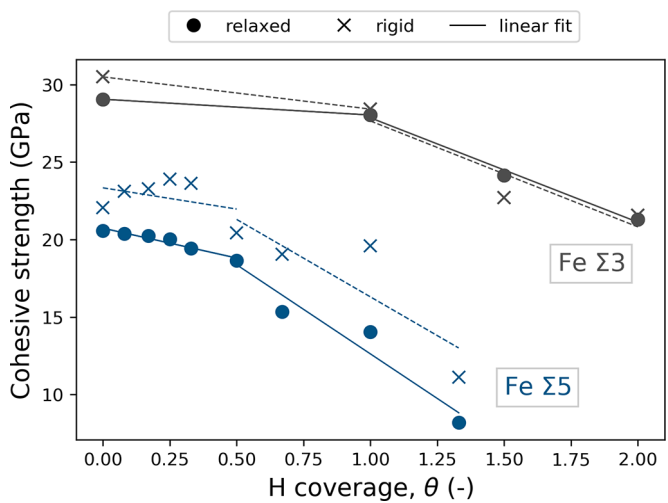

FIG. 5. Cohesive strength (= peak value of the cohesive stress) with respect to the H coverage, obtained via different schemes (relaxed and rigid calculations) for Σ5 and Σ3 STGBs.

Although the difference between the rigid and relaxed tensile tests can be up to 6 GPa, the general trends identified in both cases coincide. There are two different regimes, related to the occupancy of H in the same GB polyhedral unit, or structural unit. Up to 0.5 H coverage, only one H atom occupies the structural unit and the reduction of strength is weak. While between 0.5 and 1.33 the H occupancy is increased, leading to a much more significant reduction in strength. Also note that even for the highest value of H coverage, the Σ3 grain boundary remains stronger than the Σ5.

### B. Hydrogen solubility at zero stress and constant chemical potential

In order to estimate if the coverage values that were used in the previous section are realistic, i.e., energetically favorable, the solution energy of the solute atoms for relaxed atomic positions were calculated according to Eq. (1), i.e., for zero opening of the cohesive zone and the chemical potential of H in a H molecule. The result is shown in Fig. 6.

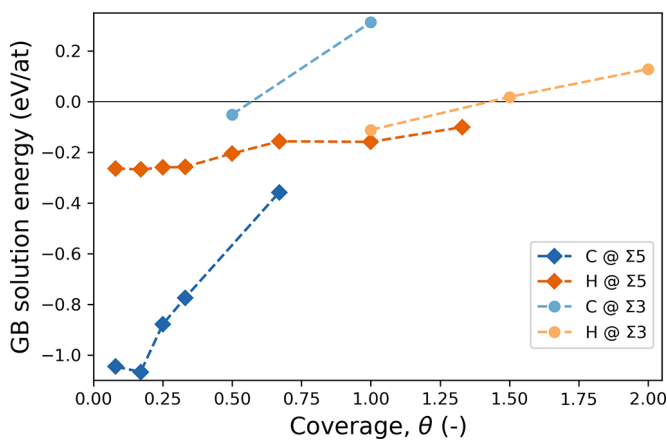

FIG. 6. GB solution energy of H and C atoms at equilibrium with varying coverage, for Σ5 and Σ3 STGBs.

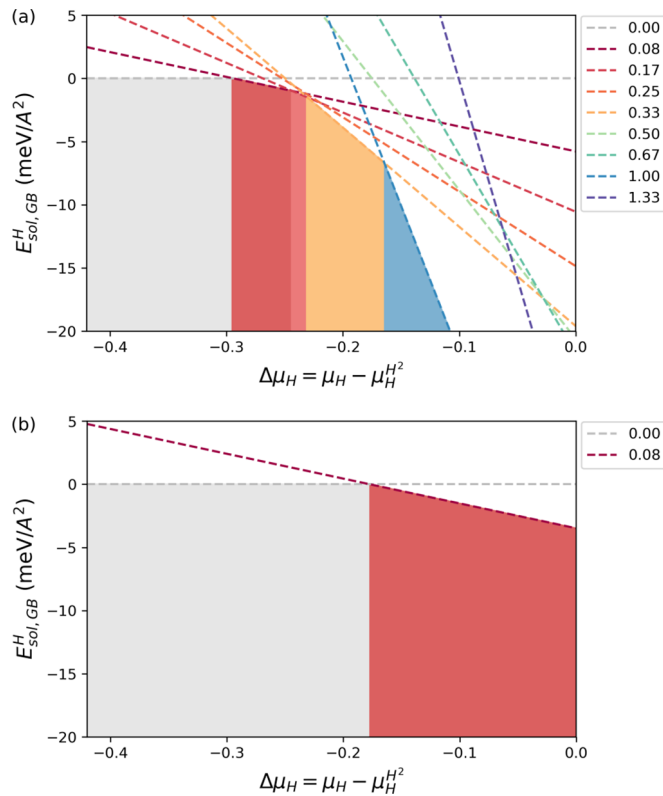

FIG. 7. Solution energy of H at the Σ5 STGB as a function of the chemical potential difference $\Delta\mu_H$ calculated at (a) 0 GPa and (b) 20 GPa. The dashed lines indicate varying values of H coverage $\theta$ at the GB, the colored regions represent the H coverage where the solution energy is the lowest.

Regarding the Σ5 GB in Fig. 6(a), it can be seen that all chosen coverage values of H and C at the GB have a negative solution energy, meaning it is more favorable for the atoms to be at the GB than at their respective reservoirs. At lower concentrations of C, the solution energy is more negative than that of H, but at increasing concentrations they become comparable. In the Σ3 case, however, due to a more closed-packed structure at the GB with less available space for segregation, when there is more than one atom in the structural unit the solubility of H is not favorable.

### C. Solution energy as function of stress and chemical potential

To estimate the effect of residual or applied stresses on the hydrogen distribution, as well as to see how much it can be influenced by varying the chemical potential of H, the solution energy is calculated as defined in the following equation:

$$E_{\text{sol},H}^{\text{GB}}(\sigma) = \frac{E_{\text{tot}}^{Fe+H}(\sigma) - E_{\text{tot}}^{Fe}(\sigma) - n_H\mu_H}{2A}. \quad (5)$$

Here, the total energies of the pure Fe and Fe+H GB systems are subtracted at the same stress, utilizing the excess energy from the first-principles cohesive zone model, defined in Eq. (2). The chemical potential, $\mu_H$ is now a variable that decreases from the value of the reference chemical potential, $\mu_H^{H_2}$, which is that of the hydrogen molecule down to $-0.4\,eV + \mu_H^{H_2}$. Figure 7(a) shows the solution energy as calculated in Eq. (5) at the Σ5 in the style of a defect phase diagram [44]. It can be seen how the H coverage increases with increasing chemical potential.

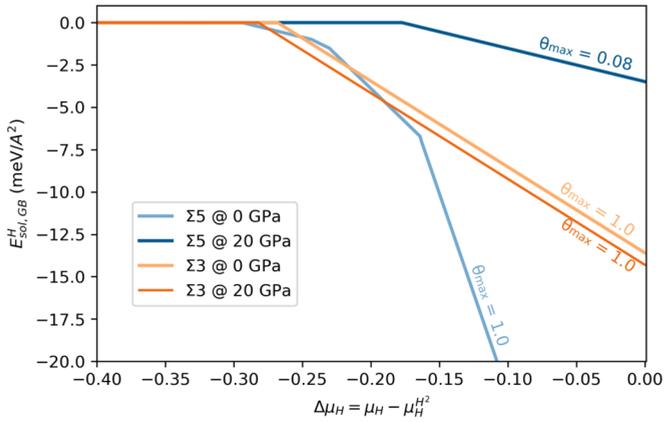

FIG. 8. Solution energy of H at the Σ5 and Σ3 STGB as a function of the chemical potential difference for selected stress states. The coverage values are the maximum values reached for the corresponding applied stress.

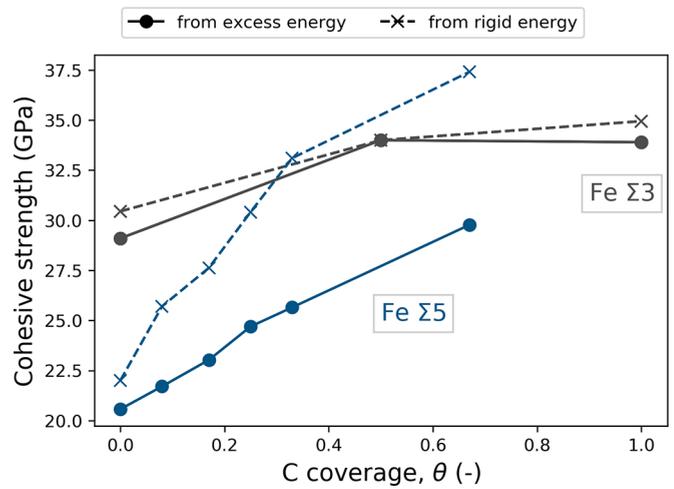

FIG. 9. Cohesive strength (= peak value of the cohesive stress) with respect to the C coverage, obtained via different schemes (from excess and rigid energies) for Σ5 and Σ3 STGBs.

We have added a third dimension to such a defect phase diagram by repeating this calculation at different values of stress. This allows us to estimate how much the two grain boundaries compete for H under mechanical loading. For demonstration, in Fig. 7(b) the case of the Σ5 STGB for 20 GPa is shown, which corresponds to the stress just before the cleavage of the GB plane occurs. The equivalent diagrams for the Σ3 STGB can be found in the Appendix B (Fig. 13).

In Fig. 8, the two boundaries and stress states are compared, it shows the resulting solution energy at the GB with respect to the chemical potential difference at different values of applied stress. In each case the concentration increases (solution energy decreases) with increasing chemical potential, indicating the sequential filling of the available segregation sites at the GB plane. Note, however, that at the Σ5 STGB, the solution energy increases, as the stress increases, while for the Σ3 STGB, we observe the opposite. Thus, the maximum coverage of the Σ5 at higher applied stress is reduced to $\theta = 0.08$, while at zero stress it reaches $\theta = 1.00$. This leads to the important phenomenon, that with increasing stress, the most favorable GB for H segregation is the Σ3 instead of the Σ5, opposite to the situation at zero stress, as observed also in Fig. 6.

### D. Cohesion enhancing effect of carbon

In the case of steels, it is necessary to study the influence of carbon to understand the embrittlement mechanisms and whether the cohesion-enhancing effect of C can counter the detrimental impact of H. Figure 9 presents the results of the calculated cohesive strength as a function of the C coverage. The strength of the Σ5 STGB increases linearly with a higher local C concentration at the GB plane. The increase from 20.6 GPa up to 29.8 GPa represents a maximum of 45%. Although the pure Fe Σ3 has a higher strength than Σ5, the cohesion enhancing effect of C is limited to a 17% increase, from 29.1 GPa to 33.9 GPa, with a negligible difference between both C coverage values studied. The resulting cohesive strength of the *ab initio* tensile test using rigid grain shifts is also reported, labeled as "rigid energy", while there is negligible difference between both procedures for the Σ3, the more open Σ5 has a considerable difference of 10 GPa. This indicates that energy relaxation is needed to fully understand the effect of segregating atoms at the GB and is crucial in the case of defects that allow more available space for segregation [17].

Another mechanism of interest in the case of ferritic steels is the C and H interaction. To understand the cosegregation effect of C and H on the cohesive properties of the Σ5 STGB, two different cases were investigated. In the first case, the supercell was constructed with a coverage $\theta = 0.33$ and the number of atoms segregating to the structural unit was limited to one. The resulting cohesive strength of the varying C and H co-segregation cases can be observed in Fig. 10(a). In the second cosegregation case studied two atoms sit in the structural unit with a total coverage $\theta = 0.67$ [Fig. 10(b)]. The cohesive strength in both cases is reduced with increasing H presence at the GB. In the first case, the cosegregation behavior falls in the expected mixing behavior at higher H and C concentrations, while in the case at higher solute coverage, the results indicate a slightly detrimental effect, where the strength of the pure Fe GB is reduced 1.6 GPa. This effect is only noticed when the full structural relaxation of the cohesive zone is allowed, when the energies are obtained rigidly, the strength reduces linearly.

### E. Effect of substitutional alloying elements

Understanding the embrittlement mechanism of H in ferritic steels requires insight into the effect of common alloying elements and impurities often found in steels. In the present paper, substitutional transition metals V, Cr, and Mn are introduced in the Fe+H Σ5 STGB system. The supercell creation and optimized configurations can be found in the study of Subramanyan *et al.* [6]. The composition of the substitutional elements in the Fe supercell corresponds to $Fe_{38}X_2$ with X = V, Cr, or Mn. The V substitutes an Fe atom sitting at the GB plane, while Cr and Mn substitute a matrix atom that sits in the layer next to the GB. The chosen coverage values of the H atom are 0.33, 0.67, and 1.33, corresponding to 1H, 2H, and 4H at the structural unit of the GB.

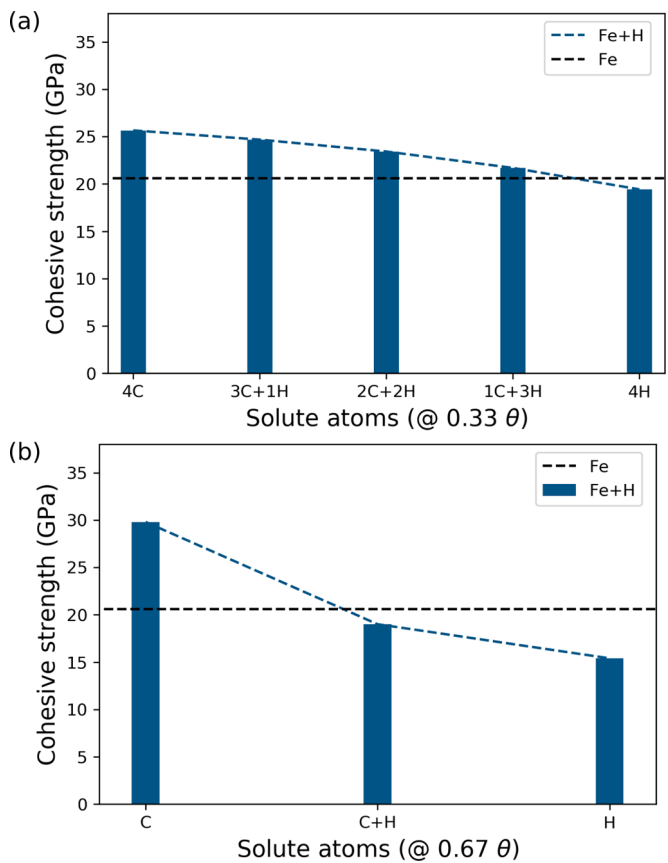

FIG. 10. Cohesive strength vs different H and C cosegregation cases for Σ5 STGB with varying H and C fraction per structural unit (SU): in (a) the maximum number of solute atoms per SU is one, total solute coverage = 0.33 and in (b) one SU contains two solute atoms, total solute coverage = 0.67.

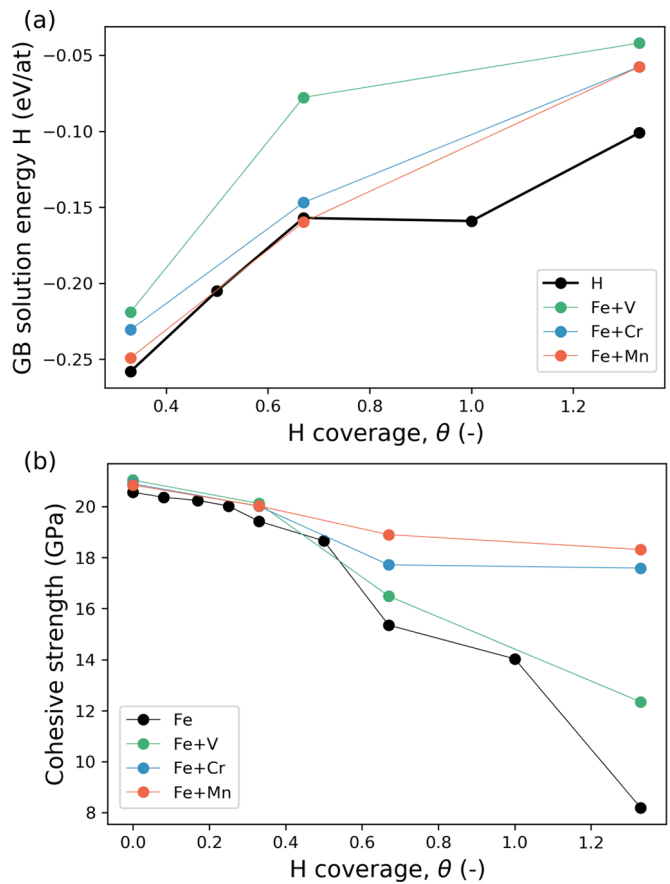

FIG. 11. (a) Solution energy of H at the GB and (b) cohesive strength vs H coverage of a GB decorated with alloying elements: V, Cr, and Mn.

The effect of the cosegregation of substitutional elements and H atoms on the solution energy of H and the cohesive strength can be observed in Fig. 11. When only one atom sits at the structural unit, the influence of the substitutional elements on the H solubility at the GB is a reduction of up to 0.04 eV/atom wrt. the pure Fe GB, as observed in [6]. However, at the highest H, coverage the effect is enhanced, reducing the solubility of H by 0.12 eV/atom, in the case of V. On the other hand, at a high local concentration of H at the GB, the strengthening effect of the substitutional atoms is considerable, vanadium increases the strength by 5.5 GPa, chromium at the GB increases the strength by 9.8 GPa; and, manganese increases the cohesive strength by 10.1 GPa.

## IV. DISCUSSION

The present study investigates hydrogen-enhanced decohesion (HEDE) in bcc Fe symmetrical tilt grain boundaries (STGBs): Σ5(310)[001] 36.9° and Σ3(112)[1$\bar{1}$0] 70.53°.

The cohesive properties of the GBs are calculated through *ab initio* tensile tests. However, stress and strain in the microstructure change not only the trapping strength of the solute atoms but also the number and arrangement of trapping sites at the GB (see energy surface of the Fe+H system in Fig. 2). Thus, a method for identifying relevant trapping sites in the structure has been proposed, based on the voids of the polyhedral unit at the GB [40]. The sites obtained through the void analysis method were compared with the sites obtained according to the energy surface and the proposed method proved to be a more efficient option. Once the trapping sites were found, the solubility of H calculated at zero stress indicated that the concentration of the solute atoms can be increased locally at the GB; considering that more than one H atom can sit in the same structural unit.

One of the key findings of this paper is that the presence of H can reduce the cohesive strength of the Σ5 GB up to 60% and Σ3 GB up to 16% (see Fig. 5). Such a strong effect has not been reported in the literature of first-principles studies of HEDE in GBs. The work of Tahir *et al.* [5] found that a monolayer of H reduced the strength 6%, at a coverage of one H atom per structural unit of the Σ5 GB. Similarly, Momida *et al.* [16] reported a 4% reduction of the Σ3(112) ideal strength. In contrast, Katzarov, and Paxton [15] calculated a decrease from 33 GPa to 22 GPa with increasing H concentration in the (111) cleavage plane in bcc Fe, using tight-binding calculations. The strong reduction of the cohesive strength at the GB plane is only observed at higher local concentrations of H when more than one solute atom sit at the structural unit.

The two grain boundaries chosen for this studies significantly differ in their atomic structure and thus in their interaction with H. The Σ5 was chosen as a sample of structures with a more open atomic environment, with a higher excess length (0.31 Å) and higher energy (1.58 J/m$^2$) than the Σ3. Meanwhile, the Σ3 is chosen as an example of a close-packed structure with a bulk-like environment and lower excess length and energy (0.16 Å and 0.43 J/m$^2$, respectively). The local atomic environment at a GB can be characterised by the bond length and excess volume. This geometric criterion has been proposed for C segregation to bcc Fe GBs [45] and for H segregation to Ni GBs [46]. This observation can now be further clarified and formalized: In this paper, the GBs are characterised in terms of the number of sites available for segregation at the GB and the size of the interstitial voids identified. The grain boundary, in which the larger number of comparatively bigger voids can be found in the relaxed structure, i.e., the Σ5 STGB, attracts hydrogen more strongly, as can be seen from the solution energies as a function of coverage, Fig. 6. However, when the grains are separated, the increase in volume leads to a decreasing solution energy at the Σ3 grain boundary, while the solution energy increases at the Σ5 GB. This is in line with the predictions of the effective medium theory, namely that there is an optimum electron density to embed H in the interstitial sites of a metal lattice [47], and the findings for H at GBs in Ni, that the smaller sites exhibit a higher electron density [48]. Thus we speculate that the structural unit at the Σ5 STGB provides an ideal volume (and electron density), while the structural unit at the Σ3 is smaller than that, but improves upon applying the separation. Accordingly, also the cohesive strength of the Σ3 GB is less sensitive to the coverage with H than that of the Σ5, although in both cases there is GB weakening due to the presence of H.

The cohesive properties of the segregated grain boundaries also strongly depend on the thermodynamic limits of the separation process. According to Hirth and Rice [22,23], these properties vary between the limit of constant composition, and of constant chemical potential. The results presented in Sec. III A correspond to constant composition, where the separation is faster than the diffusion of the H atoms. Contrarily, the limit of constant chemical potential refers to a slower separation of the interface, where some segregation sites can be filled, thus changing the composition. In Sec. III C, the solution energy is calculated under mechanical load to elucidate the effect of the chemical potential on the cohesive properties. The comparison shows another key finding of this paper: The solubility of H at the Σ3 STGB is significantly less sensitive to both, changes in the chemical potential and an increase of the stress state than at the Σ5. In the latter case, it decreases drastically as the cohesive strength of 20 GPa of the grain boundary is approached. Also, there is a cross over of solubility energies between Σ5 and Σ3 in Fig. 8, with an increasing advantage for the Σ3 as the chemical potential is increased. This means that for a suitably chosen $\mu_H$, the whole system can be stabilized by driving the H towards the Σ3 as the stress increases. This is further enabled by the fact that the energy barriers at the Σ5 GB decrease with increasing separation, as can be seen in Fig. 2, which means that the traps become more shallow. Of course, considering only two grain boundaries as potential traps is a very artificial scenario, but the study demonstrates how the stresses in the microstructure and the chemical potential of H can completely change the picture, which is obtained for DFT calculations at constant coverage, see Fig. 5. These results can inform mesoscale simulations using models, which couple hydrogen transport/diffusion equations with cohesive zone models, e.g., [49,50].

If the fracture is fast, in the sense that H will not redistribute, a high coverage at the weaker grain boundary and significant embrittlement must be expected. Thus, another important factor to consider is how the cohesion of the GB can be stabilized by adding alloying elements. The first element that comes to mind when talking about steel is C, but also the substitutional atoms V, Cr, and Mn are interesting candidates. The interplay of these elements has been investigated already in [6] but at rather low concentrations of H. However, as we know now, that coverage range underestimates the effect of segregation. Thus, in the study at hand, the range of coverage values studied was increased.

We started with the segregation of C, as well as the cosegregation of C and H, where both solutes compete for interstitial sites at the GB plane. Carbon is known to have a cohesion-enhancing effect on the Σ5 STGB, as reported in previous works [5,6,17]. In the present study, C was found to increase the strength of the Σ5 GB from 20.6 GPa to 29.8 GPa (up to 45%) and that of the Σ3 GB from 29 GPa to 33.9 GPa (up to 17%). The cohesion-enhancing effect of C is limited at higher concentrations of C at the GB, carbide formation was not considered in this study.

Another aspect that was investigated was whether the presence of C in the Fe+H system is able to counteract the detrimental effect of H on the cohesive properties of the GB (see Fig. 10). Two cases of the cosegregation of C and H were selected, based on the total solute coverage. In neither of the cases the presence of C was able to overcome the negative effect of the H segregation.

A different story is the cosegregation of H with substitutional alloying elements commonly present in ferritic steel, such as V, Cr, and Mn. The influence of these elements in the cohesive strength of the Σ5 STGB is limited at lower coverage of H ($\theta = 0$ and 0.33), and only when the H coverage increases up to 1.33 is a considerable effect observed. At the highest H coverage, the strength of the pure Fe GB is reduced by 60%, when Cr and Mn decorate the GB the reduction of the strength is 13.8% and 12.1%, respectively. Mn is found to be the alloying element, which leads to the least reduction of strength in the presence of H. This finding is in agreement with Khanchandani and Gault [51], where an atom probe tomography study on twinning-induced plasticity (TWIP) steels found increased HEDE in Mn-depleted GBs. On the other hand, the presence of V at the GB signifies a strength 35% lower than the pure Fe GB, having an overall beneficial effect considering the solubility of H is −0.05 eV/atom, higher than Cr and Mn. Although the impact of these substitutional elements at the GB does not completely negate the detrimental effect of H at the Fe STGB, the cosegregation of such elements significantly counteract the influence of H. Thus, it is necessary to discern the role of the H interaction with

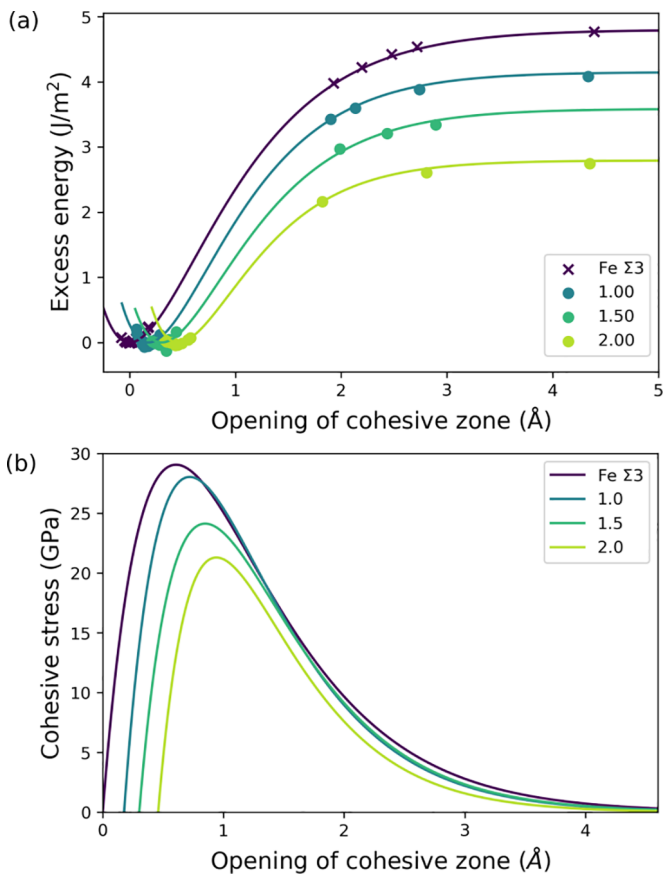

FIG. 12. (a) Excess energy and (b) cohesive stress as a function of the opening of the cohesive zone for $\Sigma 3$ STGB with varying H coverage at the GB.

TABLE I. Coverage, atoms per structural unit (SU) and areal concentration at the $\Sigma 5$ and $\Sigma 3$ STGBs.

| STGB | Coverage | Atom/SU | Areal concentration (at/Å$^2$) |
|---|---|---|---|
| $\Sigma 5$ | 0.08 | | 0.02 |
| | 0.17 | | 0.04 |
| | 0.25 | | 0.06 |
| | 0.33 | 1 | 0.08 |
| | 0.50 | | 0.12 |
| | 0.67 | 2 | 0.16 |
| | 1.00 | 3 | 0.24 |
| | 1.33 | 4 | 0.31 |
| | 1.67 | 5 | 0.39 |
| $\Sigma 3$ | 1.0 | 1 | 0.10 |
| | 1.5 | | 0.15 |
| | 2.0 | 2 | 0.20 |

The hydrogen solution energy, and thus the expected H coverage at a specific grain boundary depends on the local stress as well as the chemical potential. This dependency is very pronounced for the $\Sigma 5$ STGB, while the $\Sigma 3$ GB is only weakly affected. As a consequence, assuming fast diffusion of H in the microstructure, there is a range of chemical potential for which H atoms should redistribute, depleting the weaker $\Sigma 5$ STGB and enriching the stronger $\Sigma 3$ STGB. This would make the effect of H less detrimental. Note that the prediction of this rather unexpected effect is only possible due to the addition of a third dimension, i.e., the stress acting at the boundary, to the defect phase diagram that describes the hydrogen solution at the boundaries.

If diffusion is slow or hindered, the H distribution at zero stress will remain and the $\Sigma 5$ STGB will fail first, at a critical stress, which can be reduced by 60% in comparison to the pure Fe STGB. To prevent HEDE at slow diffusion levels, alloying elements can help. In the case of cosegregation of C and H to the GB, C exhibited cohesion-enhancing properties, although the effect is limited at higher local concentration. However, the positive influence of the segregation of substitutional elements (Cr, V, and Mn) to the $\Sigma 5$ GB on the cohesive strength of the GB plane is much more pronounced, and especially Mn can counteract the effect of H completely.

This case study shows the importance of studying HEDE at grain boundaries at higher local hydrogen coverage and finite stress values. To be able to predict the concentration, the solution energy as a function of stress, chemical potential, and coverage. Only then, the proper recipes to prevent HEDE via alloying strategies and/or by manipulating the chemical potential of H can be developed.

other alloying elements at the GB, especially at higher local concentrations of H, where a higher impact is expected.

## V. CONCLUSIONS

First-principles calculations have been carried out to investigate the segregation of H and C to bcc Fe symmetrical tilt grain boundaries: $\Sigma 5(310)[001]$ and $\Sigma 3(112)[1\bar{1}0]$. In order to identify the trapping sites at the GB a method has been implemented, based on the polyhedral unit model. This method is compared with the solution energy surfaces of H at the $\Sigma 5$ GB under strain and it proves accurate and more efficient as a site identification technique. Employing this method allowed the increase of the local H concentration at the GB to further understand HE at Fe GBs.

The hydrogen-enhanced decohesion (HEDE) mechanism is observed in both studied GBs. The $\Sigma 3$ GB was observed to have a lower susceptibility to H embrittlement, as opposed to the $\Sigma 5$ GB. This observation is associated with the GB geometry, where the more open local atomic of the $\Sigma 5$ GB translates to a detrimental effect on the cohesive properties by the presence of H. Future works could consider a formalization of the relationship between the GB geometry and the cohesive properties, specifically the GB geometry in terms of the available space for the segregation of H atoms.

## ACKNOWLEDGMENTS

This research has been supported by the Deutsche Forschungsgemeinschaft (DFG), Projects No. 414750139 and No. 535248809. The authors acknowledge the computer resources provided by the Center for Interface-Dominated High Performance Materials (ZGH, Ruhr-Universität Bochum) and the Interdisciplinary Center for Advanced Materials Simulation (ICAMS, Ruhr-Universität Bochum).

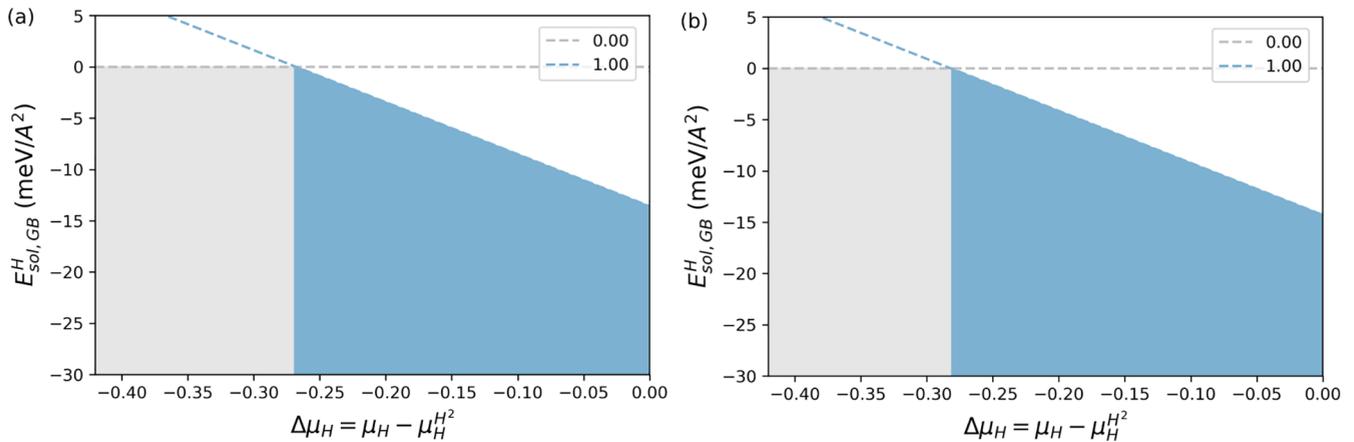

FIG. 13. Solution energy of H at the Σ3 STGB as a function of the chemical potential difference, $\Delta\mu_H$. The dashed lines (and corresponding colored area) indicate different values of H coverage $\theta$ at the GB. The applied stress is (a) 0 GPa and (b) 20 GPa.

## APPENDIX A: H COVERAGE AT THE GRAIN BOUNDARY

The coverage of the interstitial sites at the GB is calculated based on the ratio of interstitial atoms with respect to the matrix Fe atom in the structural unit that defines the GB. The corresponding areal concentration values of each configuration considered in this study are presented in Table I.

## APPENDIX B: DECOHESION OF THE Σ3 STGB

The excess energy as a function of the opening of the cohesive zone, calculated using Eqs. (2) and (3) is shown for the Σ3 STGB in Fig. 12(a). From the derivative of the UBER fit w.r.t. the opening, the cohesive stress curves shown in Fig. 12(b) are obtained. According to Fig. 6, the coverage $\theta = 2.0$ has a positive solution energy at zero stress.

The solution energy curves in Fig. 7 are produced in a similar fashion as the defect phase diagrams construction from *ab initio* calculations [44]. The calculation of the defect phase diagram concept aims to understand a defect phase and the transition between phases upon changes in the (local) chemical potential. We have added the stress in the microstructure as a third dimension. The two exemplary cases of zero and 20 GPa applied stress for the Σ3 STGB are shown in Fig. 13.